\documentclass[runningheads]{llncs}
\usepackage{graphicx}
\usepackage{amsmath,amssymb}
\begin{document}
% =================================================================
\title{Stochastic Gradient Descent Works Really Well for Stress Minimization}
\titlerunning{SGD works well}
\author{Katharina B\"orsig
\and Ulrik Brandes\orcidID{0000-0002-1520-0430}
\and Barna Pasztor}
\authorrunning{K.~B\"orsig, U.~Brandes and B.~Pasztor}
\institute{ETH Z\"urich, Z\"urich, Switzerland\\\email{ubrandes@ethz.ch}}
% -----------------------------------------------------------------
\maketitle
% -----------------------------------------------------------------
\begin{abstract}
Stress minimization is among the best studied force-directed graph layout methods
because it reliably yields high-quality layouts.
It thus comes as a surprise that a novel approach based on stochastic gradient descent 
(Zheng, Pawar and Goodman, TVCG 2019)
is claimed to improve on state-of-the-art approaches based on majorization.  
We present experimental evidence
that the new approach does not actually yield better layouts,
but that it is still to be preferred because it is simpler
and robust against poor initialization.

\end{abstract}
% -----------------------------------------------------------------

\section{Introduction}\label{sec:intro}
% =================================================================

The class of force-directed graph drawing algorithms is large
both in terms of objectives and optimization algorithms~\cite{k-fdda-13,b-fdgd-14}. 
Experimental~\cite{bp-esdbgd-09} and anecdotal evidence suggest
that a most desirable objective is
the stress function of distance-based multidimensional scaling~\cite{m-maed-66}.
Given a simple undirected graph $G=(V,E)$,
the layout $x=(\mathbb{R}^2)^V$ of a straight-line drawing is considered suitable,
if the weighted deviation
\begin{equation}
  \mathop{stress}(x)=\sum_{i<j} d_{ij}^{-2}(\|x_i-x_j\|-d_{ij})^2
  \label{eq:stress}
\end{equation}
of Euclidean distances $\|x_i-x_j\|$ in the layout
from shortest-path distances $d_{ij}$ in the graph 
is small. 

The stress function has been varied in numerous ways
to accommodate additional objectives or constraints~\cite{bm-qcsma-12,bp-mfrl-11,ghn-msmgl-13,dkm-cglsmgp-09,nob-uhmcsw-15}.
Since stress minimization is computationally intractable, 
similarly many approaches have been proposed to save computation time~\cite{imo-gmmdsgpu-09,mns-dlgmmso-18,okb-ssm-17}.
These methods are generally designed to improve an initial layout iteratively
and thus yield local minima of the stress function
that cannot be improved further by moving single vertices.

Here we are interested in assessing a recent proposal
by Zheng, Pawar, and Goodman~\cite{zpg-gdsgd-18}
that is based on stochastic gradient descent
and claimed to outperform majorization approaches~\cite{gkn-gdsm-05}.

Our own computational experiments suggest
that the new approach does not lead to better layouts,
but that it is still preferable due to its simplicity
and, crucially, indifference to initialization.
We do not address actual running times
because any comparison would be relative to the choice of speed-up techniques 
and the overall similarity of the computation suggests
that the same algorithm engineering techniques could be used in either approach.

The remainder is organized as follows. 
In Sect.~\ref{sec:techniques},
we briefly describe the proposal of Zheng et al.\ in the context of previous approaches.
The results of our experiments are presented and discussed in Sect.~\ref{sec:experiments},
and we conclude with some general implications in Sect.~\ref{sec:concl}.

\section{Stress Minimization}\label{sec:techniques}
% =================================================================

We very briefly review some major developments
in the use of multidimensional scaling in graph drawing.
This is not to provide the details of each method
but to contrast the approach based on stochastic gradient descent
with previous approaches.

\paragraph{Gradient descent.}
While first uses of multidimensional scaling for graph drawing date back to the 1960s,
it was popularized by Kamada and Kawai~\cite{kk-adgug-89},
who also introduced a localized version of the gradient descent approach used until then.
Since a necessary condition for a local minimum of the stress function is that
all partial derivatives are zero, they iteratively pick a vertex for which
the vector of partial derivatives with respect to its two coordinates has maximum length.
Then a two-dimensional Newton-Raphson method is applied to the stress function with all other vertices fixed.
Their layout is thus obtained by iteratively moving one vertex at a time
toward a position where the different stress terms cancel each other out.

\paragraph{Majorization.}
Ganser, Koren, and North~\cite{gkn-gdsm-05} proposed to use majorization~\cite{l-acamds-77} instead.
Here, the complex stress function is replaced with a convex function
that is larger for each layout but the current, for which it is equal. 
Minimizing this function leads to a new layout
that is guaranteed to have lower stress,
and the process is iterated until it converges to a local minimum.

The process can also be localized to move only a single vertex
such that the majorizing function is reduced. 
This yields an intuitive algorithm because the update 
$$
  x_i\gets \frac{1}{\sum\limits_{j\neq i}d_{ij}^{-2}}
             \sum_{j\neq i} d_{ij}^{-2}\cdot\frac{x_j+d_{ij}(x_i-x_j)}{\|x_i-x_j\|}
$$
places vertex~$i$ directly into a position
that balances out the influences of all other vertices.
One iteration consists of an update of each vertex.

Because of its simplicity and guaranteed convergence,
this approach is considered the state of the art.

\paragraph{Stochastic gradient descent.}
In this method the gradient is replaced by an unbiased estimator.
For additive objective functions such as the stress function in Eq.~\eqref{eq:stress},
the estimator may simply be a single term of the sum.
Since stress has one term for every pair of vertices, 
the contribution of this term can be reduced by
moving the two vertices either closer together or farther apart.

A single update thus moves both vertices along the vector~$\delta$  
to extend or shrink the line segment $\overline{x_i x_j}$ to match the target length~$d_{ij}$ more closely,
$$
 \begin{array}[c]{l}
   x_i \gets x_i-\frac{\mu(t)}{2}\cdot\delta\\[1ex]
   x_j \gets x_j+\frac{\mu(t)}{2}\cdot\delta
 \end{array}\qquad\text{where}\qquad
 \delta = \frac{\|x_i-x_j\|-d_{ij}}{\|x_i-x_j\|}\cdot(x_i-x_j)~, 
$$
and~$\mu(t)=\min\{1,d_{ij}^{-2}\eta(t)\}$ is a weighted step width capped at~1. 
Since an individual move is almost certainly in conflict
with the desired distances of other pairs,
the method does not converge in general.
Instead, the unweighted step width~$\eta(t)$
is made to exhibit an exponential decay over iteration time~$t$,
and convergence is thus enforced.

% -----------------------------------------------------------------
\begin{figure}
    \mbox{\includegraphics[width=0.25\linewidth]{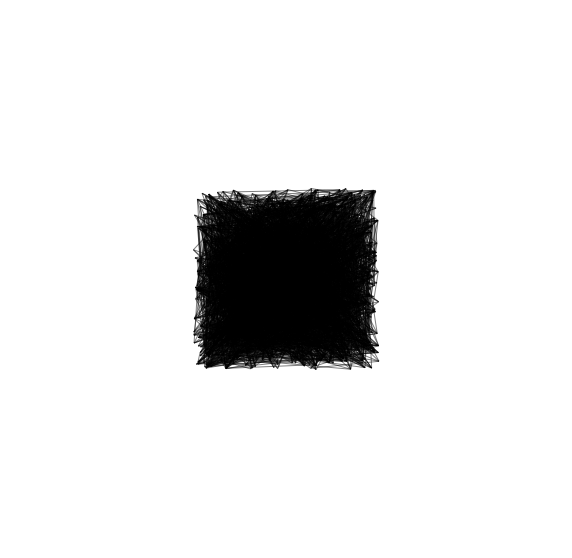}%
    \includegraphics[width=0.25\linewidth]{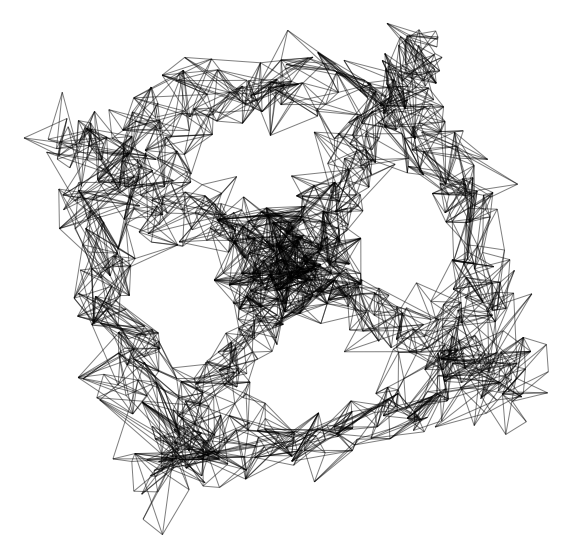}%
    \includegraphics[width=0.25\linewidth]{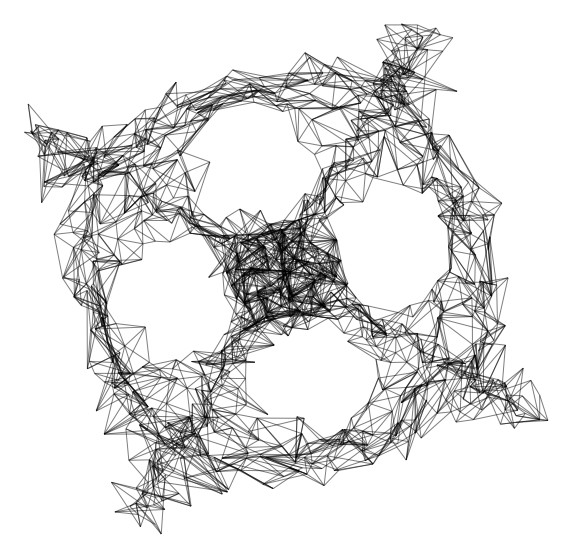}%
    \includegraphics[width=0.25\linewidth]{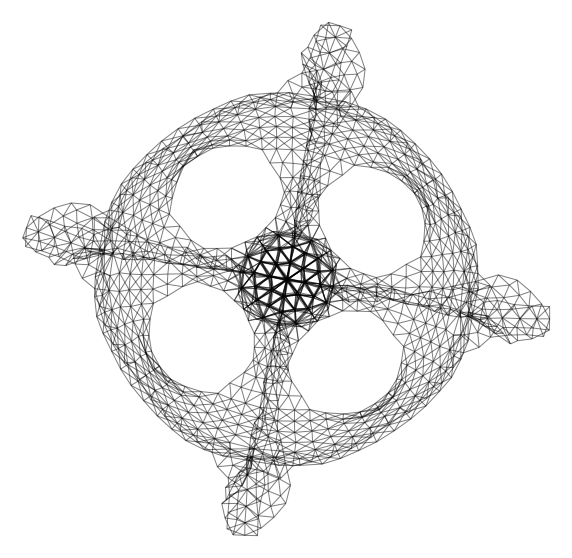}}
    \caption{Example run of stochastic gradient descent on graph \texttt{dwt\_1005} with random initialization and intermediate layouts after 1, 6, and 15 iteration.}
    \label{fig:example}
\end{figure}
% -----------------------------------------------------------------

One iteration consists of an update of all pairs of vertices in random order.
The method is thus similar to localized majorization
but instead of aggregating the influence of all other vertices before moving one,
those influences are considered separately in random order. 
The running time of one iteration is in $\Theta(n^2)$
for both stochastic gradient descent and localized majorization, 
but instead of over a linear number of linear-time vertex movements
the computation is spread out over a quadratic number of constant-time dyadic updates.

% Jakobsen 2001 (dyadic updates)
% Dwyer 2009 (capping of movement)

\section{Experiments}\label{sec:experiments}
% =================================================================

Our experiments address the claim~\cite{zpg-gdsgd-18}
that stochastic gradient descent~(SGD) outperforms majorization~(SMACOF).
The graphs used as benchmarks 
are from the University of Florida sparse matrix collection~\cite{dh-ufsmc-11}.

\paragraph{On par, but not better.}
The claim of superior performance is based on experiments 
in which both approaches are initialized with a random layout
as in the example in Fig.\ref{fig:example}. 
It was already concluded from earlier experiments, however,
that the performance of majorization depends on the initialization
and that random initialization leads to poor local minima~\cite{bp-esdbgd-09}.

% -----------------------------------------------------------------
\begin{figure}
    \centering
    \begin{tabular}{r|cccl}
    initially & 1 iteration & 6 iterations & 15 iterations\\\hline
    \raisebox{0.1\linewidth}{random}
    & \includegraphics[width=0.27\linewidth]{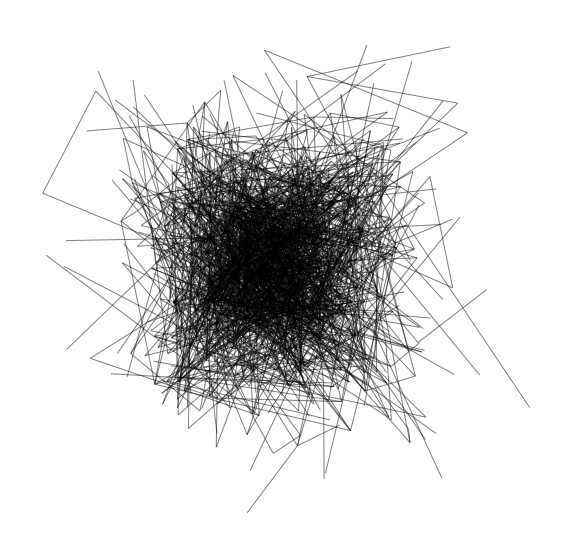}
    & \includegraphics[width=0.27\linewidth]{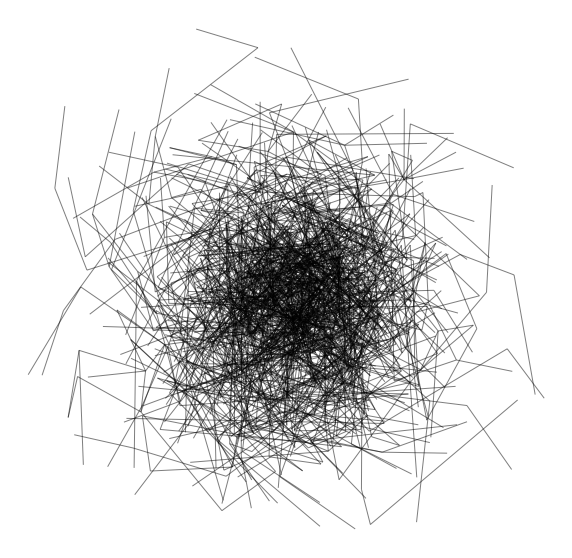}
    & \includegraphics[width=0.27\linewidth]{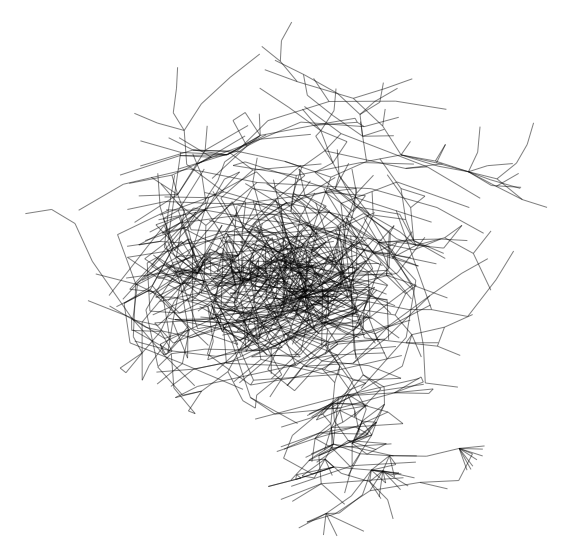} 
    & \raisebox{0.1\linewidth}{SMACOF}\\
    \raisebox{0.1\linewidth}{random} 
    & \includegraphics[width=0.27\linewidth]{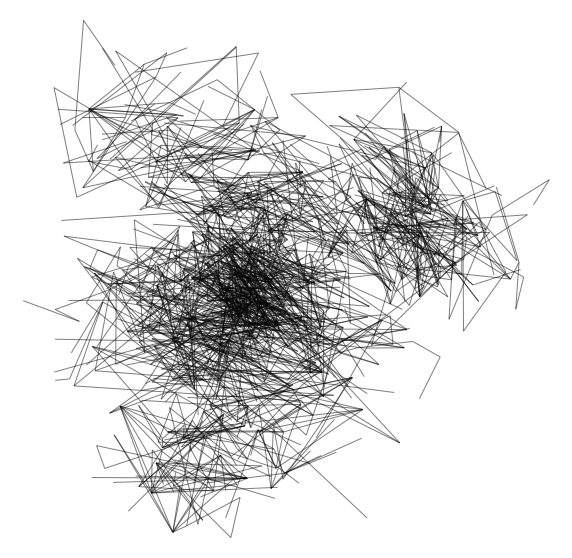}
    & \includegraphics[width=0.27\linewidth]{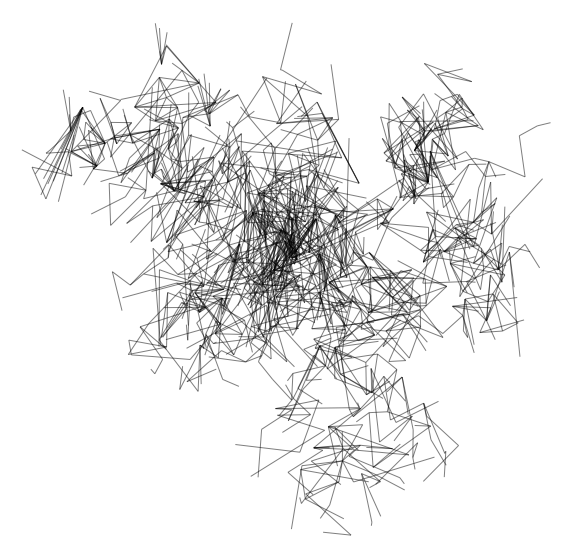}
    & \includegraphics[width=0.27\linewidth]{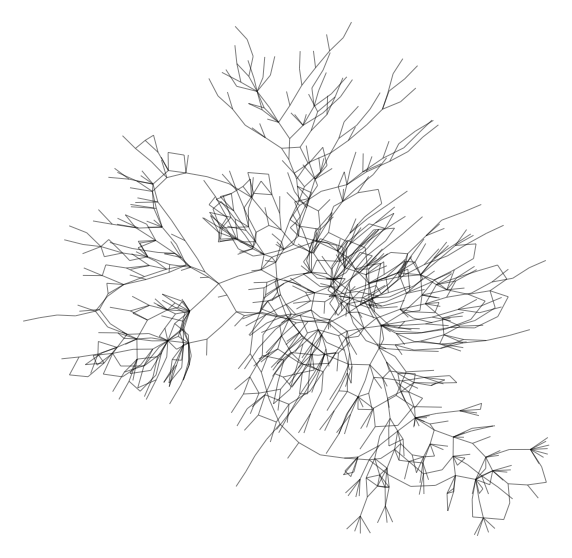} 
    & \raisebox{0.1\linewidth}{SGD}\\
    \raisebox{0.1\linewidth}{CMDS} 
    & \includegraphics[width=0.27\linewidth]{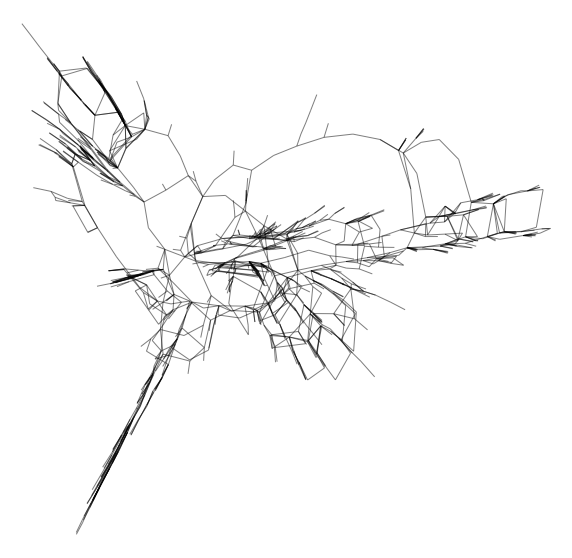}
    & \includegraphics[width=0.27\linewidth]{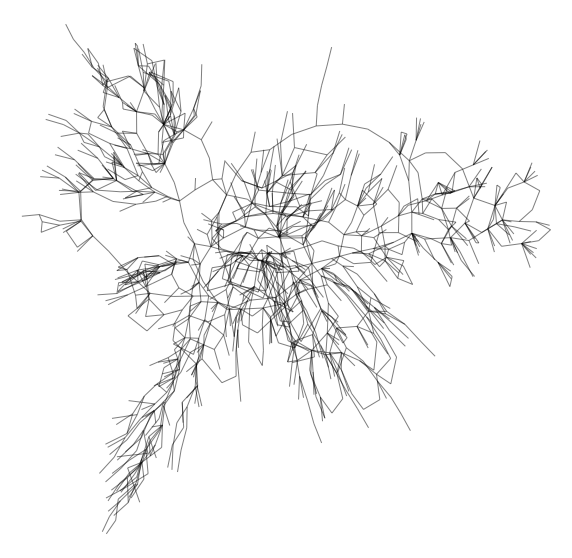}
    & \includegraphics[width=0.27\linewidth]{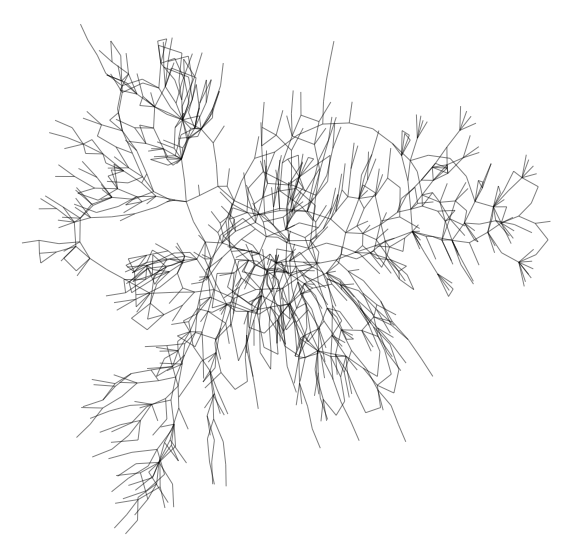} 
    & \raisebox{0.1\linewidth}{SMACOF}\\
    \raisebox{0.1\linewidth}{CMDS} 
    & \includegraphics[width=0.27\linewidth]{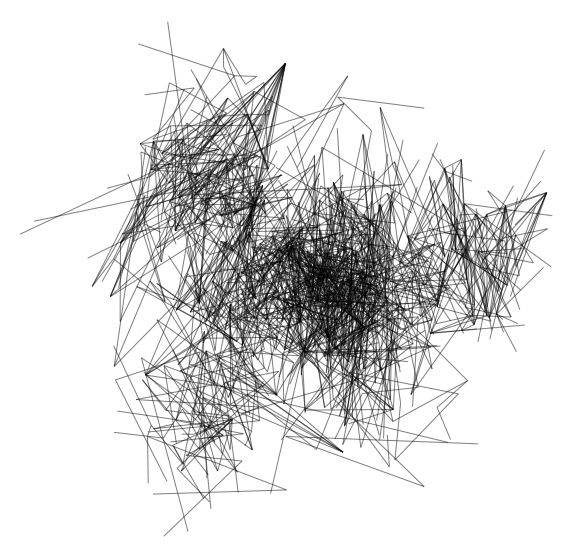}
    & \includegraphics[width=0.27\linewidth]{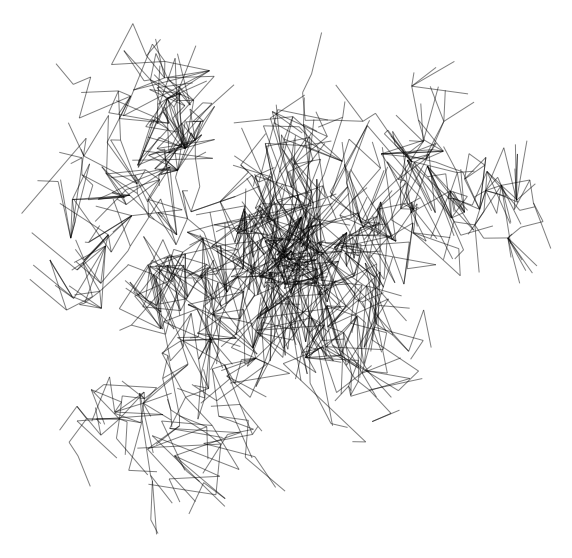}
    & \includegraphics[width=0.27\linewidth]{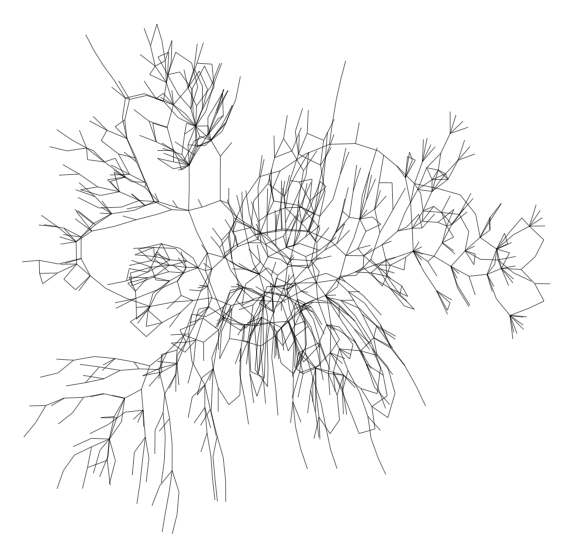} 
    & \raisebox{0.1\linewidth}{SGD}
    \end{tabular}
    \caption{An example graph (\texttt{1138\_bus}) after 1, 6, and 15 iterations.}
    \label{fig:layouts}
\end{figure}
% -----------------------------------------------------------------

% -----------------------------------------------------------------
\begin{figure}[ht]
\includegraphics[width=\textwidth]{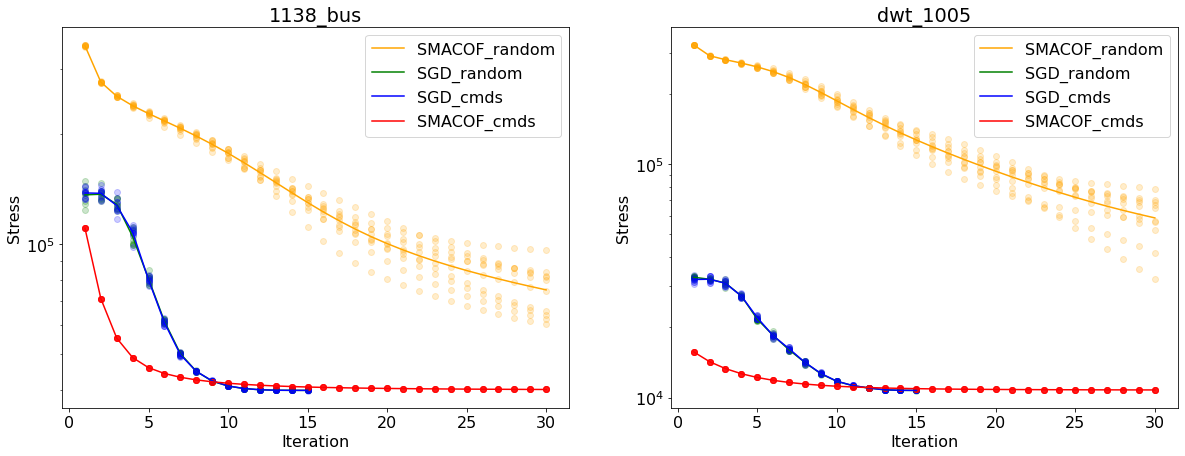}
\caption{Stress values for SGD and SMACOF on two example graphs. Random initialization is within a unit square whilst classical MDS is used at an appropriate scale. The plots show results of 10~runs for each algorithm, with circles representing single runs and lines interpolating through the means of all 10~runs. Initial stress omitted.}
\label{fig:stress}
\end{figure}
% -----------------------------------------------------------------

% -----------------------------------------------------------------
\begin{figure}[b!]
\includegraphics[width=\textwidth]{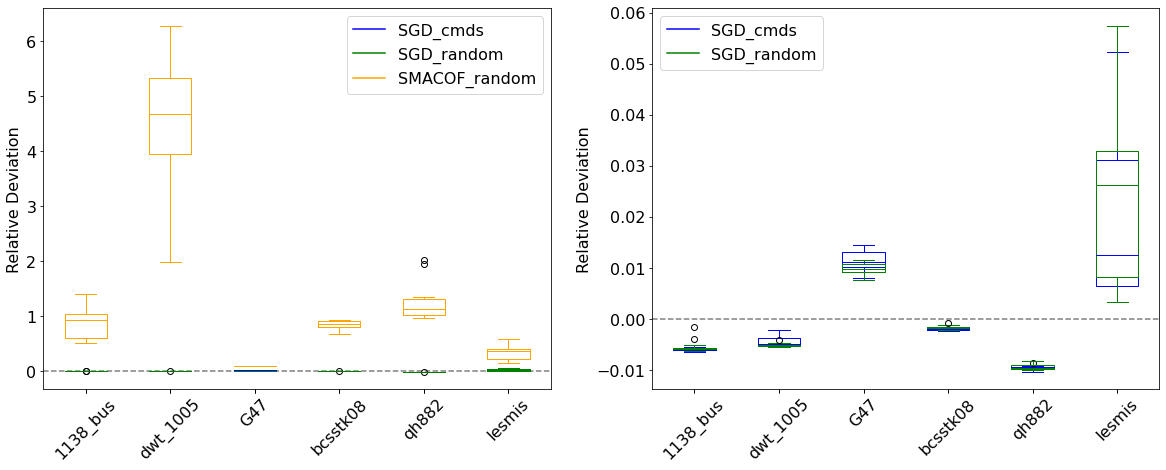}
\caption{Stress relative to baseline from SMACOF after CMDS.
With 10~runs for each instance, we find that 
random initialization results in significantly higher stress for SMACOF (left chart). 
The stress obtained from SGD differs by about $\pm1\%$ (rescaled on the right).}
\label{fig:deviation}
\end{figure}
% -----------------------------------------------------------------

We therefore ran experiments comparing the reduction in stress 
when initializing at random or with classical MDS~(CMDS).
Classical MDS results in layouts that are essentially unique and represent large distances well.
Moreover, it can be approximated at comparatively negligible cost using PivotMDS~\cite{bp-empms-07}.
Two typical examples of the results are shown in Fig.~\ref{fig:stress},
and for a better intuition, 
we also show some of the corresponding layouts in Fig.~\ref{fig:layouts}.
While the result on all benchmark graphs confirm 
that SGD indeed yields much lower stress than majorization 
when initialized with a random layout,
there is no noteworthy difference in the final stress 
when the initial layout takes care of the global arrangement.
Notably, the result of SGD is largely independent of the initialization strategy. 

Our experiments on a much larger set of benchmark graphs support these conclusions.
The evaluations in Fig.~\ref{fig:deviation} confirm quantitatively 
that majorization with random initialization is a poor baseline 
because it results in significantly higher stress 
compared to majorization after classical scaling. 
Whether SGD or the latter combination yield lower stress depends on the graph,
but relative differences are small, anyway.

\paragraph{Self-initializing.}
The seeming indifference of SGD to the initial layout prompted a second suite of experiments.

We hypothesized that the initially large displacements in SGD 
are responsible for the overall quality of the final outcome. 
If this was the case, then the differences between SGD and SMACOF should 
disappear when we initialize SMACOF with a small number of SGD iterations. 

% -----------------------------------------------------------------
\begin{figure}[tb]
\includegraphics[width=\textwidth]{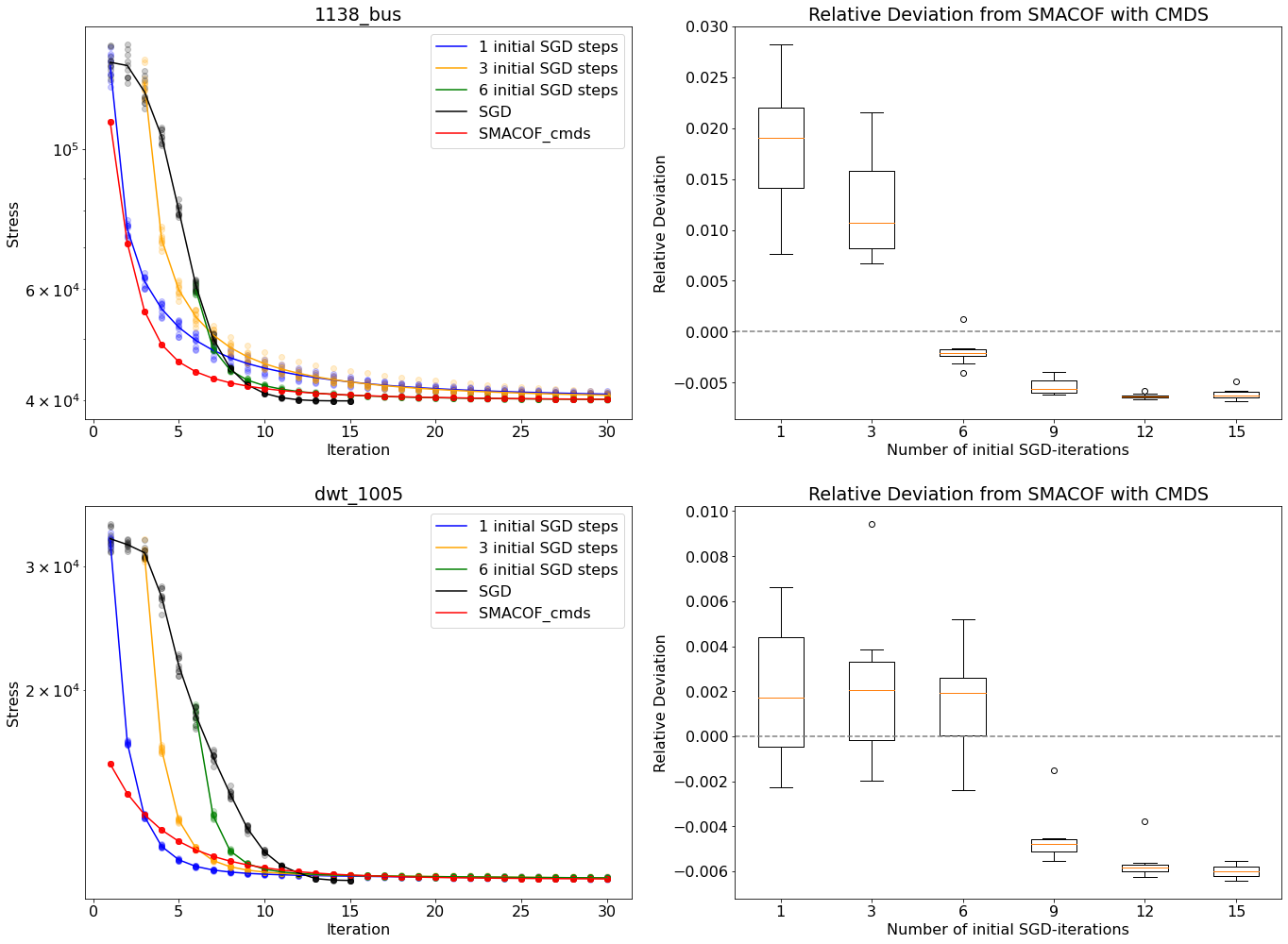}
\caption{Stress reduction by SMACOF initialized with CMDS or a few iterations of SGD (left) 
and the relative deviation of the final stress from the baseline of SMACOF with CMDS (right)
on example graphs \texttt{1138\_bus} and \texttt{dwt\_1005}. 
The initial iterations of SGD start from a random initialization in the unit square,
and each instance was run 10~times.} 
\label{fig:initialization}
\end{figure}
% -----------------------------------------------------------------

As illustrated in Fig.~\ref{fig:initialization} this is indeed the case.
Even a single step of SGD prevents majorization from sinking into a poor local minimum.
After about seven iterations of SGD, majorization yields layouts 
that are even slightly better than those obtained from initialization with CMDS.
We also note that in the next iterations, 
SMACOF reduces stress faster than SGD,
but the number of iterations to the final layout is roughly the same for both.
This number becomes smaller than for SMACOF initialized with CMDS,
offsetting the higher cost of SGD iterations compared to PivotMDS. 

We conclude that a, if not the, major advantage of the approach based on stochastic gradient descent
is the reliable untangling of any initial layout during the first few iterations.
No separate initialization strategy is required.

\paragraph{Well designed.}
We performed a number of additional experiments 
that generally confirm the recommendations given for stochastic gradient descent~\cite{zpg-gdsgd-18},
and indicate that little can be gained by straightforward attempts at improvement
such as an initial focus on long distances or the integration of majorization steps.

\section{Conclusions}\label{sec:concl}
% =================================================================

We have presented computational experiments comparing two approaches 
for graph drawing by multidimensional scaling of shortest-path distances.

Contrary to claims by the authors, we do not find that stochastic gradient descent, 
which was recently proposed as an alternative to majorization,
leads to better layouts~\cite{zpg-gdsgd-18}.
We find no significant differences in stress,
provided that majorization is initialized appropriately. 

The true advantage of stochastic gradient descent
appears to lie in its indifference to initialization. 
It is striking that this very simple and uniform algorithm
yields results that are on par with the state of the art.

We did not compare running times in this short paper because
both approaches largely perform the same operations in different order
and speed-up techniques such as subsampling and spatial aggregation abound.
Since many of these apply similarly to both approaches, 
we expect differences to be too subtle for any general claims.
Since pairs in a maximal matching can be updated without interference,
stochastic gradient descent appears to be more amenable to parallelization, though.

% difficult to incorporate soft anchoring constraints or similar into SGD? smaller initial step width...

% =================================================================
% \bibliographystyle{splncs04}
% \bibliography{references}

\begin{thebibliography}{10}
\providecommand{\url}[1]{\texttt{#1}}
\providecommand{\urlprefix}{URL }
\providecommand{\doi}[1]{https://doi.org/#1}

\bibitem{b-fdgd-14}
Brandes, U.: Force-directed graph drawing. In: Kao, M.Y. (ed.) Encyclopedia of
  Algorithms, pp.~1--6. Springer-Verlag (2014).
  \doi{10.1007/978-3-642-27848-8\_648-1}

\bibitem{bm-qcsma-12}
Brandes, U., Mader, M.: A quantitative comparison of stress-minimization
  approaches for offline dynamic graph drawing. In: Proceedings of the 19th
  International Symposium on Graph Drawing (GD~2011). Lecture Notes in Computer
  Science, vol.~7034, pp. 99--110. Springer-Verlag (2012).
  \doi{10.1007/978-3-642-25878-7\_11}

\bibitem{bp-empms-07}
Brandes, U., Pich, C.: Eigensolver methods for progressive multidimensional
  scaling of large data. In: Kaufmann, M., Wagner, D. (eds.) Proceedings of the
  14th International Symposium on Graph Drawing (GD'06). Lecture Notes in
  Computer Science, vol.~4372, pp. 42--53. Springer-Verlag (2007).
  \doi{10.1007/978-3-540-70904-6\_6}

\bibitem{bp-esdbgd-09}
Brandes, U., Pich, C.: An experimental study on distance-based graph drawing.
  In: Proceedings of the 16th International Symposium on Graph Drawing (GD'08).
  Lecture Notes in Computer Science, vol.~5417, pp. 218--229. Springer-Verlag
  (2009). \doi{10.1007/978-3-642-00219-9\_21}

\bibitem{bp-mfrl-11}
Brandes, U., Pich, C.: More flexible radial layout. Journal of Graph Algorithms
  and Applications  \textbf{15}(1),  157--173 (2011). \doi{10.7155/jgaa.00221}

\bibitem{dh-ufsmc-11}
Davis, T.A., Hu, Y.: The university of florida sparse matrix collection. ACM
  Transactions on Mathematical Software  \textbf{38}(1), ~1 (2011).
  \doi{10.1145/2049662.2049663}

\bibitem{l-acamds-77}
De~Leeuw, J.: Applications of convex analysis to multidimensional scaling. In:
  Barra, J.R., Brodeau, F., Romier, G., Van~Cutsem, B. (eds.) Recent
  Developments in Statistics, pp. 133--145. North Holland Publishing Company
  (1977),
  \url{http://www.stat.ucla.edu/\~{}deleeuw/janspubs/1977/chapters/deleeuw\_C\_77.pdf}

\bibitem{dkm-cglsmgp-09}
Dwyer, T., Koren, Y., Marriott, K.: Constrained graph layout by stress
  majorization and gradient projection. Discrete Mathematics  \textbf{309}(7),
  1895--1908 (2009). \doi{10.1016/j.disc.2007.12.103}

\bibitem{ghn-msmgl-13}
Gansner, E.R., Hu, Y., North, S.C.: A maxent-stress model for graph layout.
  IEEE Transactions on Visualization and Computer Graphics  \textbf{19}(6),
  927--940 (2013). \doi{10.1109/TVCG.2012.299}

\bibitem{gkn-gdsm-05}
Gansner, E.R., Koren, Y., North, S.C.: Graph drawing by stress majorization.
  In: Proceedings of the 12th International Symposium on Graph Drawing (GD'04).
  Lecture Notes in Computer Science, vol.~3383, pp. 239--250. Springer-Verlag
  (2005). \doi{10.1007/978-3-540-31843-9\_25}

\bibitem{imo-gmmdsgpu-09}
Ingram, S., Munzner, T., Olano, M.: Glimmer: Multilevel {MDS} on the {GPU}.
  IEEE Transactions on Visualization and Computer Graphics  \textbf{15}(2),
  249--261 (2009). \doi{10.1109/TVCG.2008.85}

\bibitem{kk-adgug-89}
Kamada, T., Kawai, S.: An algorithm for drawing general undirected graphs.
  Information Processing Letters  \textbf{31},  7--15 (1989).
  \doi{10.1016/0020-0190(89)90102-6}

\bibitem{k-fdda-13}
Kobourov, S.G.: Force-directed drawing algorithms. In: Tamassia, R. (ed.)
  Handbook of Graph Drawing and Visualization, pp. 383--408. CRC Press (2013)

\bibitem{m-maed-66}
McGee, V.E.: The multidimensional scaling of {`elastic'} distances. British
  Journal of Mathematical and Statistical Psychology  \textbf{19}(2),  181--196
  (1966). \doi{10.1111/j.2044-8317.1966.tb00367.x}

\bibitem{mns-dlgmmso-18}
Meyerhenke, H., N{\"o}llenburg, M., Schulz, C.: Drawing large graphs by
  multilevel maxent-stress optimization. IEEE Transactions on Visualization and
  Computer Graphics  \textbf{24}(5),  1814--1827 (2018).
  \doi{10.1109/TVCG.2017.2689016}

\bibitem{nob-uhmcsw-15}
Nocaj, A., Ortmann, M., Brandes, U.: Untangling the hairballs of
  multi-centered, small-world online social media networks. Journal of Graph
  Algorithms and Applications  \textbf{19}(2),  595--618 (2015).
  \doi{10.7155/jgaa.00370}

\bibitem{okb-ssm-17}
Ortmann, M., Klimenta, M., Brandes, U.: A sparse stress model. Journal of Graph
  Algorithms and Applications  \textbf{21}(5),  791--821 (2017).
  \doi{10.7155/jgaa.00440}

\bibitem{zpg-gdsgd-18}
Zheng, J.X., Pawar, S., Goodman, D.F.M.: Graph drawing by stochastic gradient
  descent. IEEE Transactions on Visualization and Computer Graphics
  \textbf{25}(9),  2738--2748 (2019). \doi{10.1109/TVCG.2018.2859997}

\end{thebibliography}

% =================================================================
\end{document}